\documentclass[%
reprint,
amsmath,amssymb,
aps,
]{revtex4-1}

\usepackage{graphicx}
\usepackage{dcolumn}
\usepackage{bm}
\usepackage{hyperref}
\usepackage{booktabs}
\usepackage{comment}
\usepackage{xcolor}
\usepackage{braket}

\newcommand{\Na}{\mathcal{N}}
\DeclareMathOperator*{\argmin}{arg\,min}

\begin{document}
	\title{Social distancing in pedestrian dynamics and its effect on disease spreading
	}
\author{Sina Sajjadi$^{1}$}
\thanks{Both authors contributed equally to this work.}
\author{Alireza Hashemi$^{1}$}
\thanks{Both authors contributed equally to this work.}
\author{Fakhteh Ghanbarnejad$^{1,2}$}
\email[]{fakhteh.ghanbarnejad@gmail.com}
\affiliation{$^1$Physics Department, Sharif University of Technology, P.O. Box 11165-9161, Tehran, Iran\\
$^2$ Institute for Theoretical Physics, Technical University of Dresden, 01062 Dresden, Germany 
}
\begin{abstract}
	
        Non-pharmaceutical measures such as social distancing, can play an important role to control an epidemic in the absence of vaccinations. In this paper, we study the impact of social distancing on epidemics for which it is executable. We use a mathematical model combining human mobility and disease spreading. For the mobility dynamics, we design an agent based model consisting of pedestrian dynamics with a novel type of force to resemble social distancing in crowded sites. For the spreading dynamics, we consider the compartmental $SIE$ dynamics plus an indirect transmission with the footprints of the infectious pedestrians being the contagion factor. We show that the increase in the intensity of social distancing has a significant effect on the exposure risk. By classifying the population into social distancing abiders and non-abiders, we conclude that the practice of social distancing, even by a minority of potentially infectious agents, results in a drastic change on the population exposure risk, but reduces the effectiveness of the protocols when practiced by the rest of the population.
        Furthermore, we observe that for contagions which the indirect transmission is more significant, the effectiveness of social distancing would be reduced. This study can provide a quantitative guideline for policy-making on exposure risk reduction.
\end{abstract}
	
	\maketitle
	
\section{\label{intro}Introduction}

	The ongoing COVID-19 pandemic has had severe consequences on nations worldwide. It has been considered as one of the costliest disasters after WWII and has imposed so many costs to the local, regional and global markets including hundreds of billions to the global insurance industry, tourism and other businesses. 
    Facing these difficulties without any approved vaccine so far, governments have turned to non-pharmaceutical measures such as social
    distancing to limit the transmission of the disease and to hinder the growth of the spreading dynamics \cite{noauthor_social_nodate}.
    Motivated by this situation, we mathematically study the \textit{effectiveness} of social distancing and how it can be implemented in order to reduce the risk of spreading.
	
	Following the early work of Kermack and McKendrick \cite{kermack_mckendrick_1927} researches have implemented compartmental models, i.e., categorizing the population into collectively exhaustive compartments, to describe and predict the epidemic dynamics.
	While initial research in the field mostly focused on mean-field approximations \cite{keeling2011modeling}, later on, a large body of work was concentrated on network studies with either analytical or computational approaches. \cite{newman_2018, barrat}.
	And then due to the greater access to agent level interaction data, the role of temporality of interactions got more attention in the studies
	\cite{masuda2017temporal,temporalnetworksholme, smallbutslowworld, fakhteh-hospital, sajjadi_impact_2020}.
		
    Mobility of agents has been studied with different approaches including  pedestrian dynamics introduced in \cite{helbing_social_1995}. While this model has been widely implemented and studied, only recently it has been used to model transmission processes. For example \cite{namilae_self-propelled_2017} and \cite{harweg_agent-based_2020} study the level of exposure of individuals to the infection, while \cite{namilae_multiscale_2017, kim_coupling_2020, gosce_analytical_2015} develop population level equations and \cite{xiao_modeling_nodate, bouchnita_multi-scale_2020, derjany_multiscale_2020} introduce agent based models to study the spreading in pedestrian dynamics.
	
	In this work we define a novel type of social distancing (keeping distance from other agents to avoid infection) based on the pedestrian dynamics.
	We also study the indirect transmission by taking into account the role of environment as a vehicle of spreading \cite{us2013principles}. We will investigate the system for different scenarios and a range of parameters. We show where social distancing is \textit{executable} and how it can be \textit{effective} for lowering the \textbf{exposure risk}, however, in some regime of parameters, the increase in indirect transmission may cancel the \textit{effectiveness} of social distancing. 
	As we will discuss, our results can be used as a quantitative guideline for policy-making on reducing the exposure risk for many contagions, including SARS-CoV-2.
	
\section{\label{model}Model}
	\subsection{\label{model_mob}Mobility of the agents}
	We simulate the walking patterns of people in a closed environment. For this purpose, we implement a formulation of the pedestrian dynamics introduced in \cite{helbing_social_1995}, including a novel type of social distancing force. Agents aim to reach randomly chosen destinations while trying to keep distance from other agents and physical barriers (e.g. walls). Although the actual Newtonian forces do not have a direct impact on the dynamics, as humans rarely have physical contact with each other/barriers while moving, taking into account the social pseudo-forces governing a person's movement patterns
	enables us to study this issue in a physical framework. This method has achieved realistic results confirmed with experimental data \cite{castellano_statistical_2009}.
	
	As depicted in Fig. \ref{fig:schematic-illustration} panel a) in this dynamics, each agent's movement is governed by three forces:
	\begin{enumerate}
		\item \textbf{Personal force $\mathbf{F}_{i}^{(pers)}$}:
		
		Each agent $i$ has a tendency to move with a velocity vector $\mathbf{v}_{i}^{0} = v_{i}^{0}\hat{\mathbf{v}}_{i}^{0}$ with $v_{i}^{0}$ being the comfortable speed and $\hat{\mathbf{v}}_{i}^{0}$ the normal vector toward its chosen destination. This tendency is expressed in Eq. \ref{eqn:pers-f} as follows:
		\begin{equation}
		\label{eqn:pers-f}
		\mathbf{F}_{i}^{(pers)} = m_{i}\frac{\mathbf{v_{i}}^{0} - \mathbf{v}_{i} } {\tau}
		\end{equation}
		Where $\mathbf{v}_{i}$ denotes the agent's current velocity, $m_i$ its mass, and $\tau$ refers to the reaction time. The values for $v_{i}^{0}$ and $\tau$ have been previously calibrated in \cite{helbing_social_1995} and will be set to $v_{i}^{0}=1.3$ and $\tau=0.5$ for our study.
		
		\item \textbf{Social force $\mathbf{F}_{ij}^{(soc)}$}:

		Every agent $i$ tries to keep distance from every other agent $j$. This tendency can be modeled by a decreasing force, function of their relative distance $r_{ij}$ with the direction of the force being the normalized vector $\hat{\mathbf{r}}_{ij}$ pointing from agent $j$ to $i$. In this research we consider the exponential force as in Eq. \ref{eqn:soc-f}, introducing $\sigma_{i}$ as a measure for agent $i$'s tendency to abide by social distancing.
		We also consider $r_c$ as a cut-off distance for the simulation purposes, i.e., agents do not exert any force on other agents further from $r_c = 3$.
		
		\begin{equation}
		\label{eqn:soc-f}
		\mathbf{F}_{ij}^{(soc)} = \kappa \sigma_{i}
		{e^{-\frac{r_{ij}}{\sigma_{i}} } \hat{\mathbf{r}}_{ij} }
		\end{equation}
		
		In real life situations, the values for $\sigma_{i}$ and $\kappa$ would be determined by the environmental properties. For example, people tend to walk closer together in a shopping mall and more further apart when taking an afternoon walk in a park.
		$\sigma_{i}$ can be used to represent the idea of social distancing in a community of agents. With a higher value for $\sigma_{i}$, agent $i$ would tend to stay further from other agents.
		For simplicity purposes we consider $\kappa$ to be a constant and set $\kappa = 7$. This conforms with the previous research on pedestrian dynamics \cite{helbing_social_1995}.

		\item \textbf{Barrier avoidance force $\mathbf{F}_{iw}^{(bar)}$}:
		Agents avoid barriers in the same manner that they avoid other agents as expressed in Eq. \ref{eqn:bar-f}, with $\sigma_w$ being the uniform tendency to keep distance from physical barriers and $\kappa_{w}$ as the force constant. In our study the only barriers would be the walls surrounding the environment.
		
		\begin{equation}
		\label{eqn:bar-f}
		\mathbf{F}_{iw}^{(bar)} =\kappa_{w} \sigma_{w}
		{e^{-\frac{r_{iw}}{\sigma_{w}} } \hat{\mathbf{r}}_{iw} }
		\end{equation}
		
	\end{enumerate}
	Although as stated, these are not Newtonian forces, we can obtain the equation of motion for each agent $i$ as expressed in Eq. \ref{eqn:motion}.
	
	\begin{equation}
	\label{eqn:motion}
	m_{i} \frac{d\mathbf{v}_i}{dt}    =
	\mathbf{F}_{i}^{(pers)} +
	\mathbf{F}_{ij}^{(soc)} + \mathbf{F}_{iw}^{(bar)}
	\end{equation}
	
	Here we consider all the agents to be of the same mass $m$, so without loss of generality, we set $m = 1$.
	Upon reaching their destination, agents are assigned with new random destinations as practiced in \cite{harweg_agent-based_2020} to display a continuous motion of the agents resembling the movement in a closed environment, e.g., a mall, office, etc. Agents are also not allowed to exceed the speed limit $v_{max} = 2$.
	
	With these assumptions, we simulate the mobility of agents in different environments and scenarios according to Equations \ref{eqn:pers-f} - \ref{eqn:motion}
	using Euler method \cite{wiki:euler-method}. These numerical simulations are similar to the methods commonly used in molecular dynamics.
	We set the simulation time step $\Delta t = 0.1$ to ensure the smooth and realistic movement of the agents.
	Agents are initially positioned randomly. To ensure realistic initial distancing between the agents, the spreading dynamics begins $40$ time steps after the mobility dynamics start.
	\\
	
	\textbf{\textit{\label{social_distancing}Social Distancing}}
	
	To go further with Eq. \ref{eqn:soc-f}, firstly we set all the $\sigma_i=\sigma$. This parameter quantifies social distancing. As we intuitively expect, higher values of $\sigma$ exhibit both faster reaction to nearby agents (stronger force) and larger perceived personal space (higher range). 
	
	To validate this intuition, we proceed by defining $\braket{\overline{L}_{i,\mathcal{N}_{i}}}$ as the ensemble average of the mean of minimum distance between agents while $L_{i,j}$ is the pairwise distances of all agents $i$ and $j$ and define 
	$\Na_{i} = \argmin_j (L_{i,j})$ as agent $i$'s nearest neighbor, and
	${L}_{i,\Na_{i}}$
	as agent $i$'s minimum distance with any other agent. $\braket{\overline{L}_{i,\mathcal{N}_{i}}}$ is depicted as an increasing function of $\sigma$ in Fig. \ref{fig:sigma_distance} top panel, therefore this average provides us with an intuition about $\sigma$ since it is a good observable quantity. 
	
	\subsection{\label{model_spread}Spreading Model}
	To model the spreading dynamics, we use a compartmental model, categorizing the population into 3 compartments, $S$ (Susceptible), $I$ (Infectious) and $E$ (Exposed). Agents in state $S$ will become $E$ upon getting into contact with the infection.
	
	The time duration agents spend in a crowded environments is relatively shorter than the latent period for most infectious diseases \cite{wiki:incubation}; therefore $E$ agents are not expected to become $I$ and infect others. $I$ agents are also not expected to recover and move to a fourth compartment $R$.
	
\begin{figure}
    	\centering
    	\includegraphics[width=\linewidth]{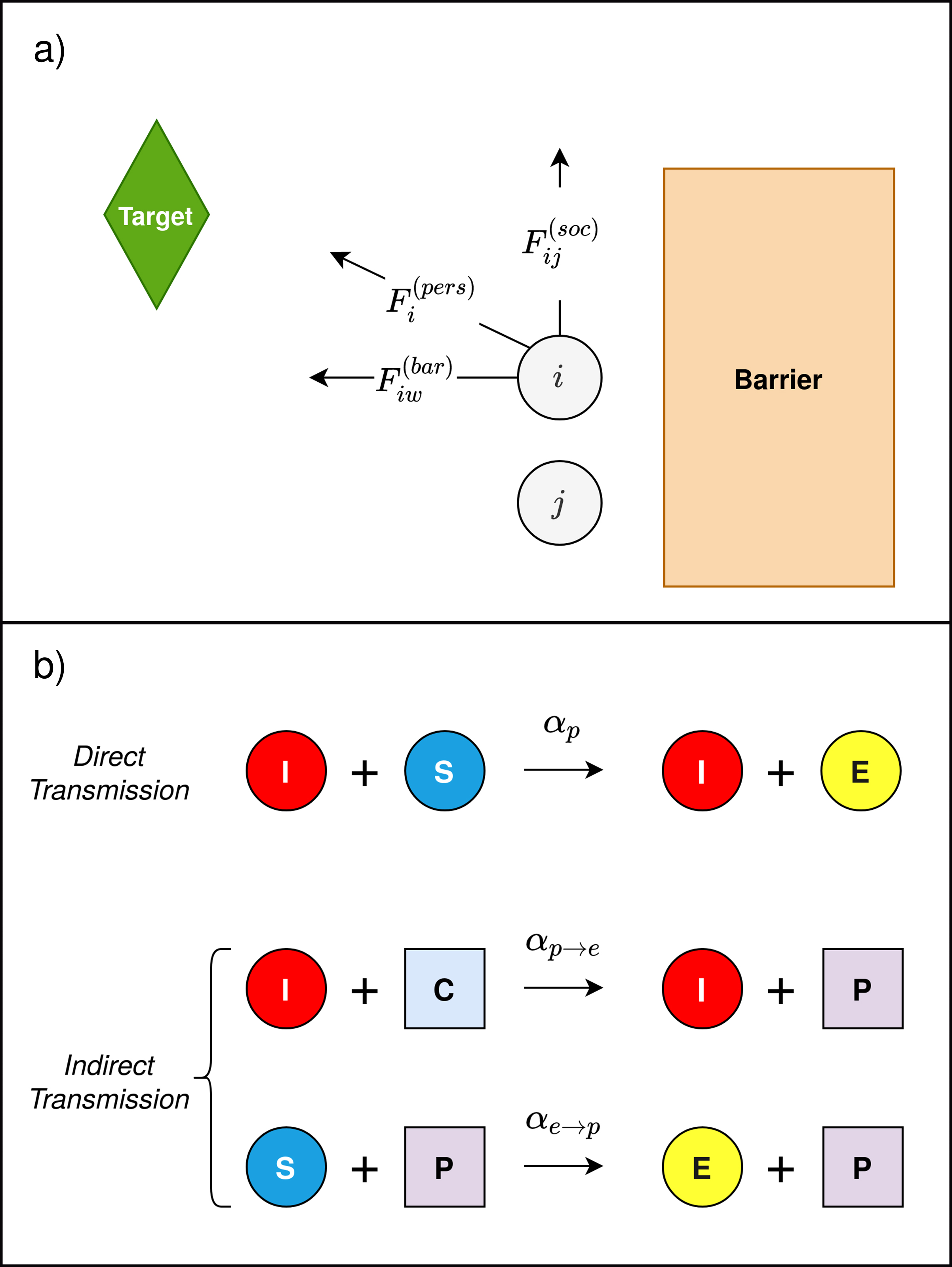}
    	\caption{A schematic illustration of the model. a) The mobility model: three types of forces exerted on agent $i$, described in
    	\ref{model_mob}. Each agent chooses a random target and moves toward it ($\mathbf{F}_{i}^{(pers)}$) while keeping distance from other agents ($\mathbf{F}_{ij}^{(soc)}$) and physical barriers ($\mathbf{F}_{iw}^{(bar)}$) (see Eq. \ref{eqn:motion}), 
    	b) The Spreading model: the two types of contagion spreading described in
    	\ref{model_spread} as direct transmission (due to person to person contact of the agents) and indirect transmission (due to the polluted environment). Circles represent agents, while $S, I, E$ denote the compartments they belong to.
    	Squares represent tiles of environment, with $C$ and $P$ respectively representing clean and polluted tiles.
    	}
    	\label{fig:schematic-illustration}
    \end{figure}

	Infections can occur in one of the following manners:
	
	\begin{enumerate}
		\item \textbf{Direct Transmission (Person-to-Person Infection)}:
		Infectious agents ($I$) infect susceptible agents ($S$) in their vicinity with the radius $r_s$, turning them to exposed $E$ agents by the probability $\alpha_{p}$ at each time step as depicted in Fig. \ref{fig:schematic-illustration} panel b).
		
		\item  \textbf{Indirect Transmission (Environmental Infection)}:
		
		Although the mobility model is formulated in a continuous manner, in order to account for the environmental pollution, 
		the environment is discretized into a lattice of size $L \times L$. Agents in state $I$ contaminate the tile they are standing on, with probability $\alpha_{p \rightarrow e}$ at each time step;
		On the other hand, $S$ agents stepping on contaminated tiles get infectious and turn $E$, with probability $\alpha_{e\rightarrow p}$
		as depicted in Fig. \ref{fig:schematic-illustration} panel b).
		For simplicity we consider $\alpha_{p\rightarrow e} = \alpha_{e \rightarrow p} =\alpha_{e}$.
		
	\end{enumerate}
	
	While in most circumstances $\alpha_p \neq \alpha_e$, the frequency of checking the possibility of infection via both methods using a rejection-based algorithm \cite{rejection-based} should be the same for the model to be consistent.

	In this study, for the sake of simplicity and to achieve more generic results, the number of initial agents in state $I$ will be always set equal to $1$ and the rest of the agents ($N-1$) will be initially in state $S$.
	
	\begin{table}[ht]
        	\caption{Model parameters and constants}
        \begin{tabular}[t]{lcc}
        $N$     & Total number of agents      \\ \hline
        $\sigma$ & Social distancing intensity ($m^{-1}$) \\ \hline
        $L=30$     & Environment size ($m$)           \\ \hline
        $\sigma_w=5$     & Barrier avoidance constant ($m^{-1}$)     \\ \hline
        $v_{i}^{0}=1.3$ & Agents' comfortable speed ($m/s$)\\ \hline
        $v_{max}=2$ & Agents' maximum speed ($m/s$)\\ \hline
        $\kappa=7$ & Social force constant ($kg.m^2.s^{-2}$) \\ \hline
        $\kappa_{w}=1$ & Barrier avoidance force constant ($kg.m^2.s^{-2}$) \\ \hline
        $r_c=3$ & Social force cutoff ($m$) \\ \hline
        $r_s=1$     & Direct infection maximum distant ($m$)           \\ \hline
        $\tau=0.5$     & Agent reaction time ($s^{-1}$)            \\ \hline \hline
        $\alpha_p$ & Direct infection probability \\ \hline
        $\alpha_e\qquad$ & Indirect infection probability \\ \hline
        \end{tabular}
        \label{table_param}
\end{table}
    A summary of model parameters is presented in table \ref{table_param} and we will explain how we fix some parameters in the next section.
    
	\section{\label{rted in esults}Results}
    We study each parameter and its effect on \textbf{exposure risk factor $E$} defined as the fraction of agents exposed to the infection. We choose parameters in ways that reflect real world scenarios.
	
	\subsection{\label{pop-dens}Population density}
\begin{figure}[]
		\begin{center}
			\includegraphics[width=\linewidth]{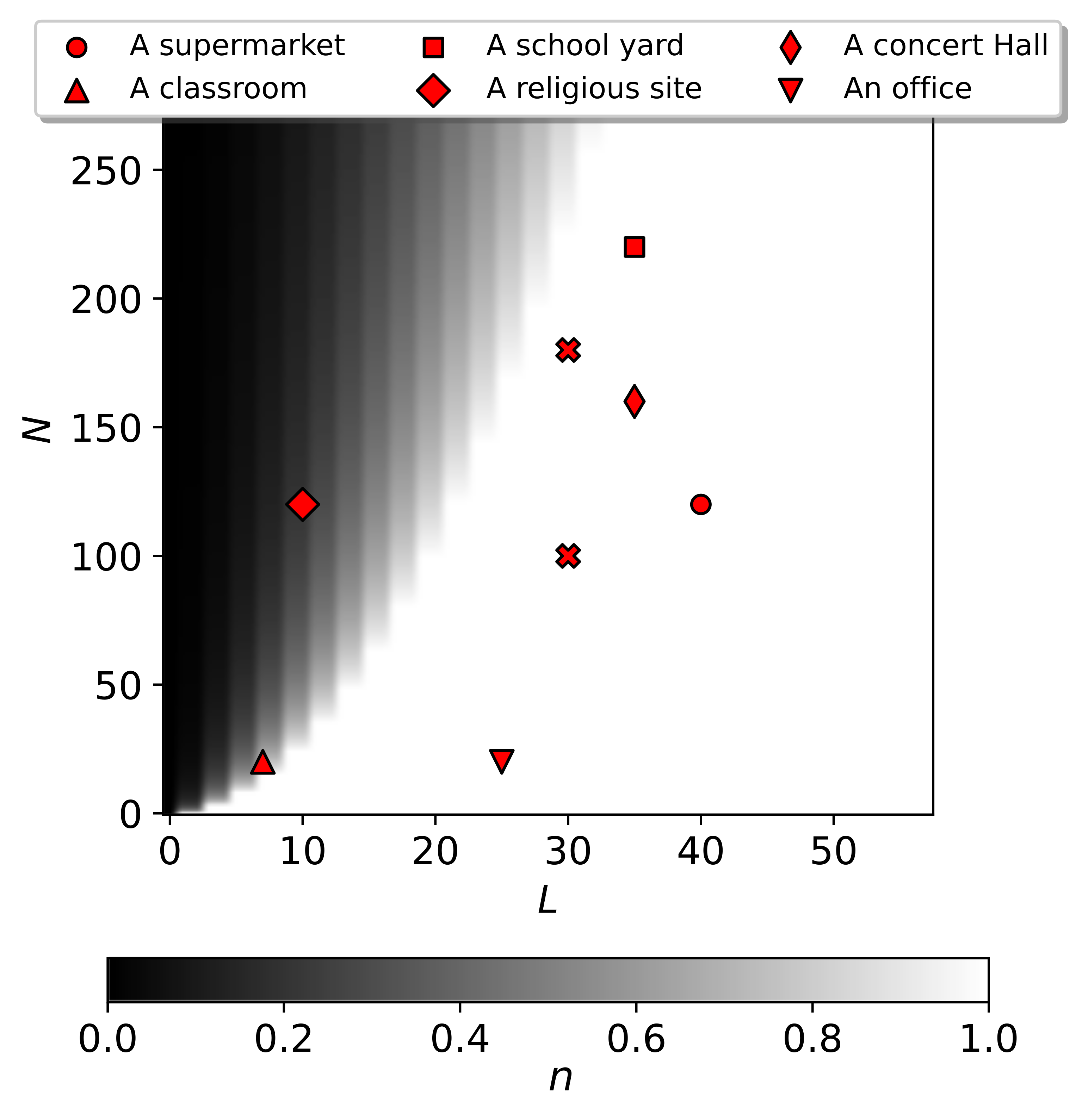}
			\caption{The proportion of the population able to maintain a 1 meter distance from each other in environments with different sizes ($L^2$). The markers correspond to empirical data from local sites. A Walmart shopping mall, a post office, the Vahdat concert hall in Tehran, Shahid Beheshti high school in Zanjan (Iran), and the Grand Mosalla mosque of Tehran (Iran) at peak hours. These data points have been normalized by different factors to only resemble the population density, not the actual size and populations. The two crosses represent the chosen data points for our simulations ($L = 30; N = 100, 180$).
			}
			\label{fig:ideal-social-distancing}
		\end{center}
	\end{figure}

	Population density is a significant factor determining the possibility of abiding by social distancing measures. In order to quantify this factor, we define social distancing limit $n$ as the proportion of the total population which are able to maintain a fixed distance from each other in a room.
	We can calculate $n$ by considering an area of $l^2$ ($l$ being the desired physical distancing) for each agent. $n = 1$ indicates a room in which total social distancing is \textit{executable} whereas $n<1$ implies otherwise. Fig. \ref{fig:ideal-social-distancing} demonstrates the curve separating two regimes $n < 1$ and $n=1$.
	Some real world locations are also depicted in our parameter space based on their size and population.
	
	For our simulation, we assume $l=r_s=1$ meter, which corresponds to the high-risk distance for COVID-19 and some other similar diseases \cite{onemeter}. We set $L=30$ and choose two different scenarios for the population density where the room is filled at $0.\overline{4}$ and $0.8$ of its the maximum capacity. This would enable us to study the \textit{effectiveness} of social distancing in the region where it is \textit{executable} ($n=1$).
	These two cases respectively refer to $N=180$ and $N=100$ and would resemble a mildly crowded and a heavily crowded environment.
	
			\begin{figure}[]
		\begin{center}
			\includegraphics[width=\linewidth]{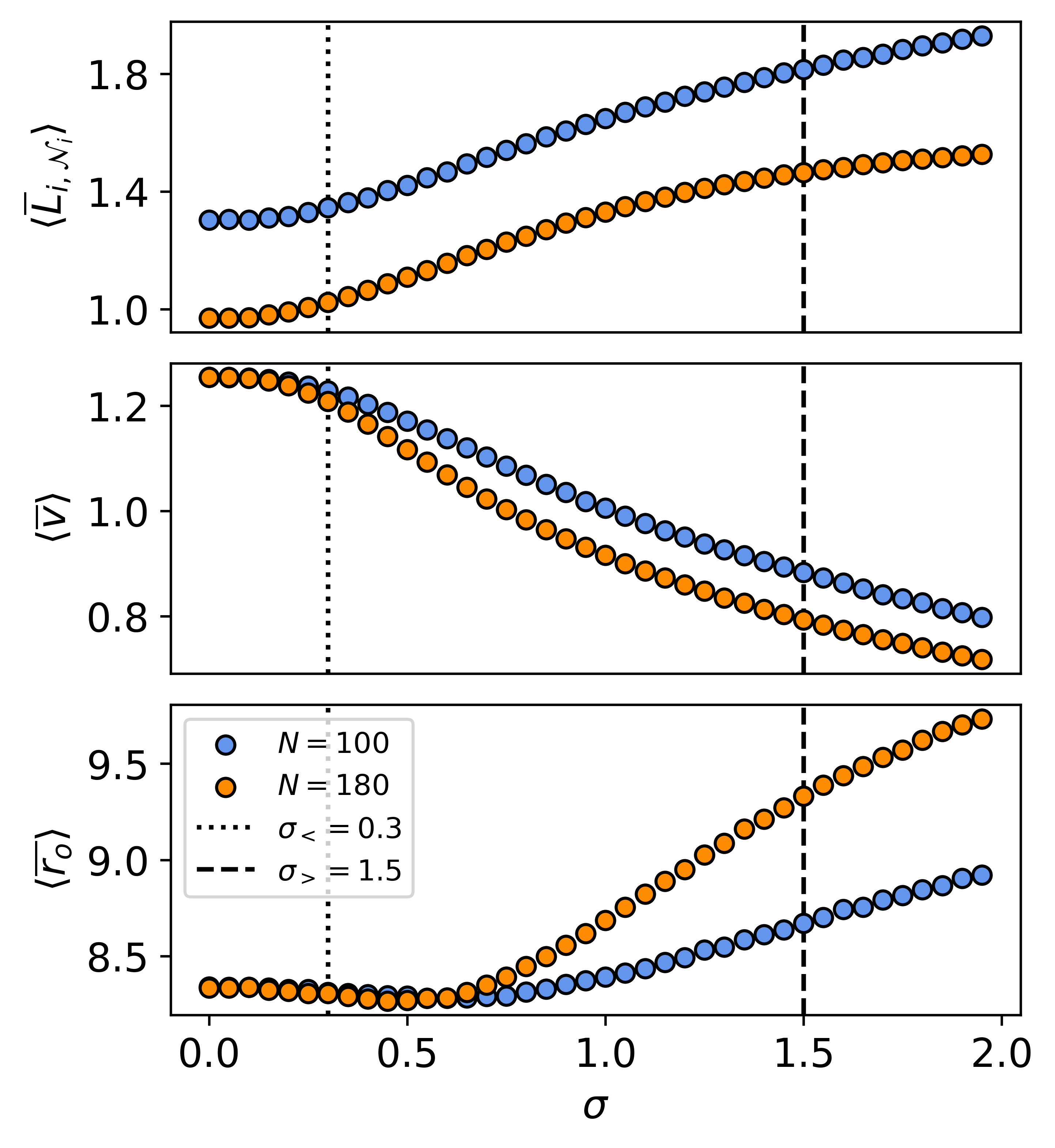}
			\caption{Mobility quantities of agents as a function of social distancing intensity $\sigma$.
			The results for two different cases of $N=100$ (blue) and $N=180$ (orange) are presented.
			Top panel: The ensemble average of mean neighbor distance $\braket{\overline{L}_{i,\mathcal{N}_{i}}}$
				as a distance-based indicator between the agents.
				Middle panel: The ensemble average of agents' speeds $\braket{\overline{v}}$
				(subject to Eq. \ref{eqn:motion}).
				Bottom panel: The ensemble average distance $ \braket{\overline{r_o}}$ from the origin for different values of social distancing intensity.
			    We observe an initial slight but significant decrease and a later increase in $\braket{\overline{r_o}}$ explained in Fig. \ref{fig:r-v}.
                The dotted and dashed lines, respectively correspond to the \textit{baseline} values of $\sigma_<$ and $\sigma_>$.
                Please note the discrepancy in the range of values for the three panels.
                \textit{Error bars are smaller than the marker size.}			}
			\label{fig:sigma_distance}
		\end{center}
	\end{figure}

	\begin{figure}[]
	\begin{center}
		\includegraphics[width=\linewidth]{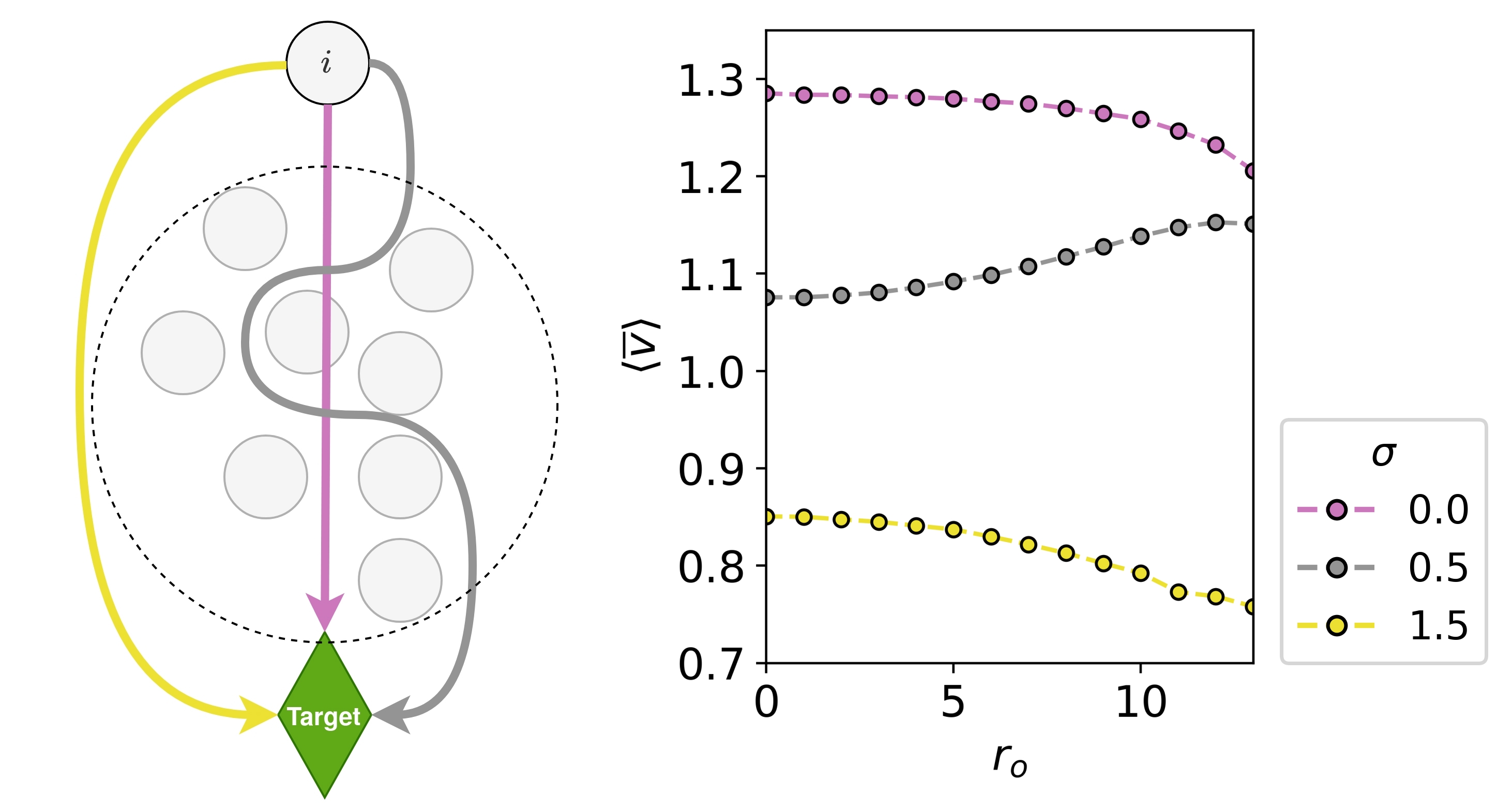}
		\caption{
	    The effect of swarm on agents' speed for $N=180$.
		Color code indicates $\sigma$ values.
		Left panel: The schematic illustration of the mobility of a single agent $i$, facing an area of high density (swarm) clustered in the center of the environment (dashed circle).
	    With $\sigma = 0$, agent $i$ moves directly toward its target without interacting with the other agents.
	    With $\sigma = 1.5$ the agent moves around the swarm, increasing $\braket{ \overline{r_o} }$ of the system compared to $\sigma = 0$.
	    With $\sigma = 0.5$ the agent moves through the swarm while making several "brakes" and minor detours from the direct path of free movement, resulting in lower speed in the central high density area.
		Right panel: The ensemble average speed for agents with $r_o$ distance from the origin. The $r_o$ values have been rounded to integers. 
		For $\sigma = 0$ and $\sigma = 1.5$, 
		$\braket{\overline{v}}|_{r_o \rightarrow 0} < \braket{\overline{v}}|_{r_o \gg 0}$
		while for $\sigma = 0.5$, 
		$\braket{\overline{v}}|_{r_o \rightarrow 0} > \braket{\overline{v}}|_{ r_o \gg 0}$.
		\textit{Error bars are smaller than the marker size.}
		}
		\label{fig:r-v}
	\end{center}
\end{figure}

	\subsection{\label{sigma}Social distancing intensity}
	As discussed earlier in section \ref{social_distancing}, we use the parameter $\sigma$ to control the intensity of social distancing. In order to get a better real-life intuition of the parameter $\sigma$, we measure the average distance which agents keep from their nearest agent while moving in the environment in this model. The results for both population densities are provided in Fig. \ref{fig:sigma_distance}.
	As mentioned, the ensemble average of agents' neighboring distances $\braket{\overline{L}_{i,\mathcal{N}_{i}}}$ monotonically increases as a function of $\sigma$ (Fig. \ref{fig:sigma_distance} top panel).
	This increase is, however, not without a cost. To stay further from other agents, each agent would have to lower their average speed and "brake" more often, resulting in lower average speed (Fig. \ref{fig:sigma_distance} middle panel) and subsequently lower efficiency of the agents in reaching their goals. Therefore the trade-off between social distancing and efficiency should be taken into account to set an intensity for social distancing.
	Near $\sigma = 0.3$ we observe that the value of $ \braket{ \overline{v} }$ for the two environments, begin to diverge.
	Based on these results we use $\sigma_<=0.3$ for the social force among agents in regular walking, as in \cite{helbing_social_1995}; and we set $\sigma_>=1.5$ for the case of social distancing applied by agents
	as an example case, in which a distance greater than $r_s$ is maintained between the agents while leaving enough room for them to move around the environment ($\braket{\overline{v}}> 0.7$).
	These choices are made without the loss of generality and in order to set parameters closer to real-life situations.
	
	In Fig. \ref{fig:sigma_distance} bottom panel, we observe $\braket{\overline{r_o}}$ the ensemble average distance from the origin, an indicator of the concentration of the agents.
	For $\sigma \rightarrow 0$, increasing $\sigma$ results in a higher condensation of agents, i.e, lower average distance from the origin while for higher values of $\sigma$ we observe the opposite effect.
	This counter-intuitive behavior can be explained considering three different scenarios of an agent interacting with a highly dense area (swarm) as illustrated in the left panel of Fig. \ref{fig:r-v}. In the $\sigma = 0$ scenario the agent moves through the swarm without losing speed, while in the $\sigma = 1.5$ case, it moves around the swarm resulting in a higher $\braket{ \overline{r_o} }$ of the system compared to $\sigma = 0$, and in the case of $\sigma = 0.5$ the agent moves through the swarm, albeit by a lower speed. This behavior results in spending more time in dense areas and effectively decreases $\braket{ \overline{r_o} }$.
	This analysis can be supported by Fig. \ref{fig:r-v} right panel. For $\sigma = 0$ and $\sigma = 1.5$ the average speed of agents located at $r_o$ distance from the origin is a decreasing function of $r_o$. This decrease is due to the fact that the center of the environment ($r_o \rightarrow 0$) acts as a corridor for reaching their targets, and agents often move through this area with a higher speed. In comparison, agents move slower when they are farther away from the center of the environment, as they brake and redirect their velocity more often.
    For $\sigma = 0.5$ the opposite is true and agents have relatively lower speed in the center of the environment, resulting in a lower $\braket{ \overline{r_o} }$ in comparison to $\sigma = 0$ and subsequently the initial decrease in Fig. \ref{fig:sigma_distance} bottom panel.
    
\begin{figure}[]
		\begin{center}
			\includegraphics[width=\linewidth]{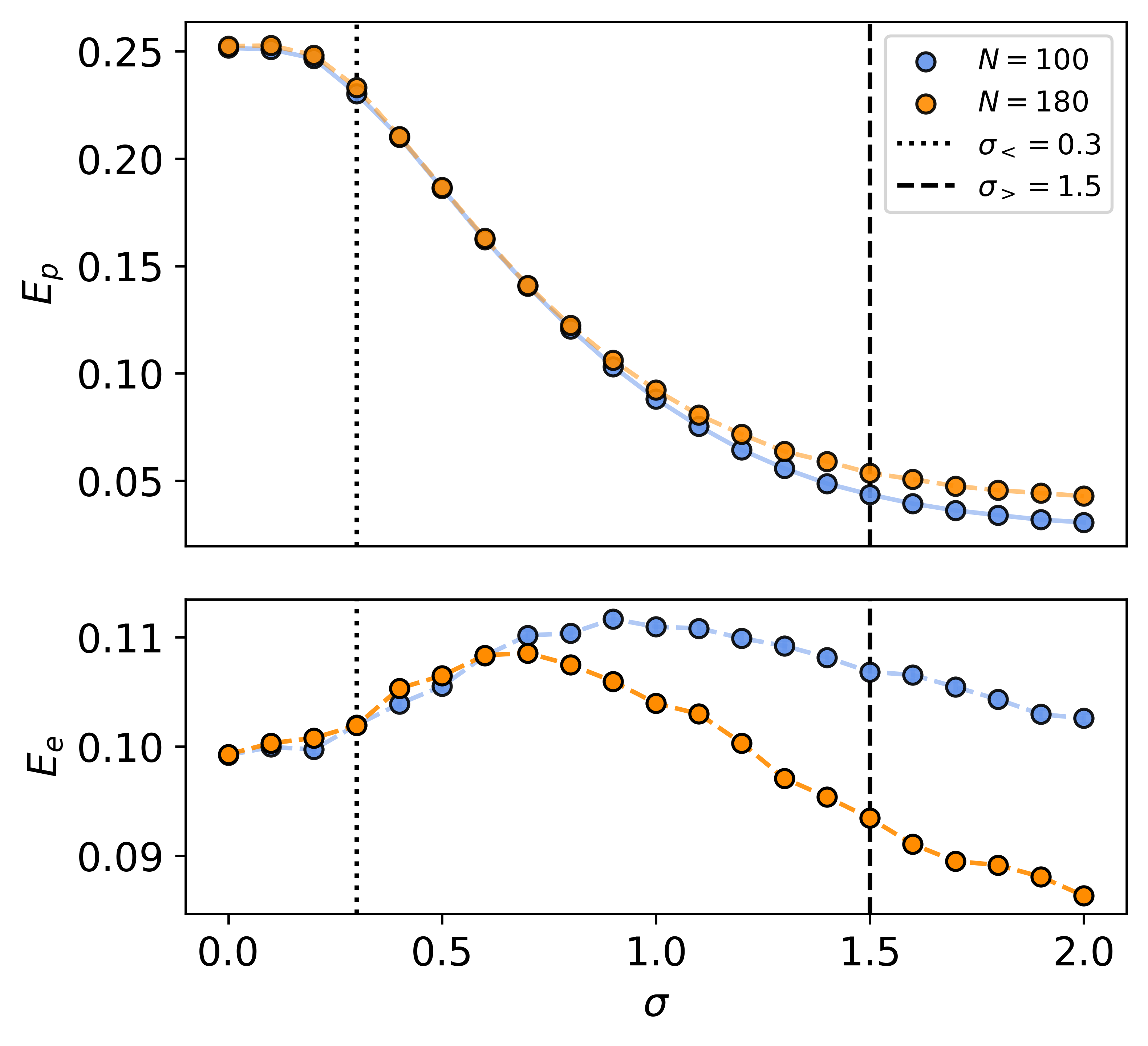}
			\caption{The average $E$
				for different values of social distancing intensity. The top panel demonstrates $E_p$ the exposure due to direct infection and the bottom panel demonstrates the exposure due to environmental infection. The results for two different cases of $N=100$ (blue) and $N=180$ (orange) are presented.
                The dotted and dashed lines, respectively correspond to the \textit{baseline} values of $\sigma_<$ and $\sigma_>$.
                Please note the discrepancy in the range of values for the two panels.
                \textit{Error bars are smaller than the marker size.}			}
			\label{fig:sigma_variaton_E}
		\end{center}
	\end{figure}

	We run the pedestrian and spreading dynamics for both population regimes discussed in \ref{pop-dens}.
	For simplicity, we choose $\alpha_e = 2 \times 10^{-3}$ and $\alpha_p = 10^{-2}$ and later on in section \ref{infection_parameters} we analyze the behavior of the model for different spreading parameters.
	In this section we set $\sigma$ uniform across all agents
	($\sigma_{i}=\sigma$).
	The average proportion of the exposed population is calculated after 10 minutes ($6 \times 10 ^ {3}$ time steps) of simulation time and plotted against the values of $\sigma$ in Fig. \ref{fig:sigma_variaton_E}.
	First observation based on this result is that for this set of parameters, the total risk, $E=E_p+E_e$, gets reduced by the increase in social distancing. For example, for $\sigma_<$ and $\sigma_>$, the value of this reduction is $E(\sigma_<)-E(\sigma_>)=0.207$ for $N=180$ and $E(\sigma_<)-E(\sigma_>)=0.181$ for $N=100$ (The effect of social distancing in other infection parameter sets will be presented in Fig. \ref{fig:two-sigmas-E-compare}). We can observe that the increase in the strength of this parameter monotonically reduces the risk of getting infected through direct infection (top panel).
	
	We can also see that the fraction of agents infected via environmental contamination slightly increases for both population regimes as we increase $\sigma$ but decreases monotonically afterwards and more intensively for $N=180$.
	This is	partly due to the \textit{competition} between the two methods of infection. The probability of a $S$ agent getting indirectly infected increases when the number of susceptible agents is higher (as a result of the weakness of the direct infection).
	Another reason for this increase and later decrease in the value of $E_e$ is the similar variations in the concentration of agents as a function of $\sigma$ (Fig. \ref{fig:sigma_distance} bottom panel).
    The higher concentration of agents in the central area of the environment
    causes higher indirect transmission. The central tiles, traversed more often by $S$ agents, will be more likely to be contaminated as they are more often walked upon by the $I$ agent.
    
	\subsection{\label{Social_distancing_commonness}Social distancing commonness}
	
\begin{figure}[]
		\begin{center}
			\includegraphics[width=\linewidth]{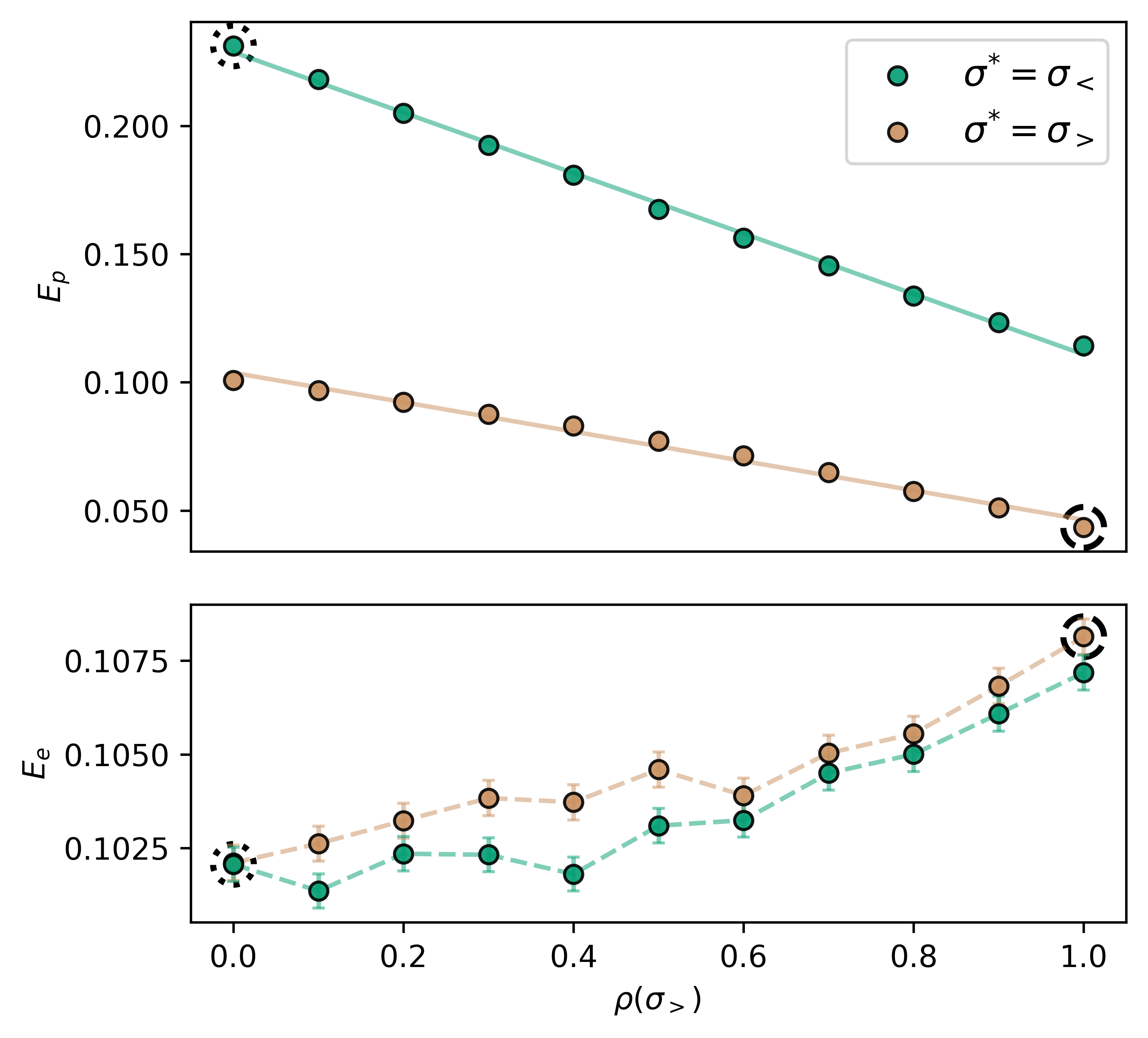}
			\caption{
				The average $E$
				for different values of social distancing intensity with $N=100$. The top panel demonstrates $E_p$ the exposure due to the direct infection and the bottom panel demonstrates the exposure due to the environmental infection. The results for two different cases where the initial infectious agent obeys social distancing ($\sigma^*=\sigma_>$, in khaki) and does not obey it ($\sigma^*=\sigma_<$, in dark green) are presented.
                The dotted and dashed circles, respectively correspond to the \textit{baseline} values of $\sigma_<$ and $\sigma_>$.
                These values signify the scenarios in which all of the population respectively ignores ($\sigma = \sigma_<$) and abides by ($\sigma = \sigma_>$) social distancing.
                Please note the discrepancy in the range of values for the two panels.
                \textit{Error bars of the top panel are smaller than the marker size.} 
			}
			\label{fig:two_type_sigma_variaton}
		\end{center}
	\end{figure}

	In this section we study the effect of the fraction of people who abide by social distancing on the total risk factor. For this goal we divide the population into two groups, a non-abiding group with $\sigma=\sigma_<=0.3$ and a social distancing group with $\sigma=\sigma_>=1.5$ and we run the simulation for different proportions of people in each group. The parameter $\rho(\sigma_>)$ would represent the proportion of the agents who abide by social distancing. We analyze this matter in two different scenarios, in which the initial infectious agent is always in the former ($\sigma^*=\sigma_<$), or the latter ($\sigma^*=\sigma_>$) group.
	The results are presented in Fig. \ref{fig:two_type_sigma_variaton} which shows the strong effect of the behavior of the initial infectious agent on the risk factor $E$. In both cases, the risk of getting exposed to the contagion drops linearly as the percent of people who abide by social distancing increases but the slope and the intercept of this linear behavior differs based on the social distancing intensity of the initial infectious agent. In other words, the risk drops by a significant factor, if only the infectious agent abides by social distancing. The risk of exposure due to indirect infection increases as more people abide by social distancing similar to the situation explained in \ref{sigma}.

	\subsection{\label{infection_parameters}Infection Parameters}
	To see the behavior of the dynamics in other regimes of the parameter space and in order to get a wider perspective of the model, we implement the simulations for various values of infection probabilities $\alpha_{p}$ and $\alpha_{e}$.
	
\begin{figure}[]
		\begin{center}
			\includegraphics[width=\linewidth]{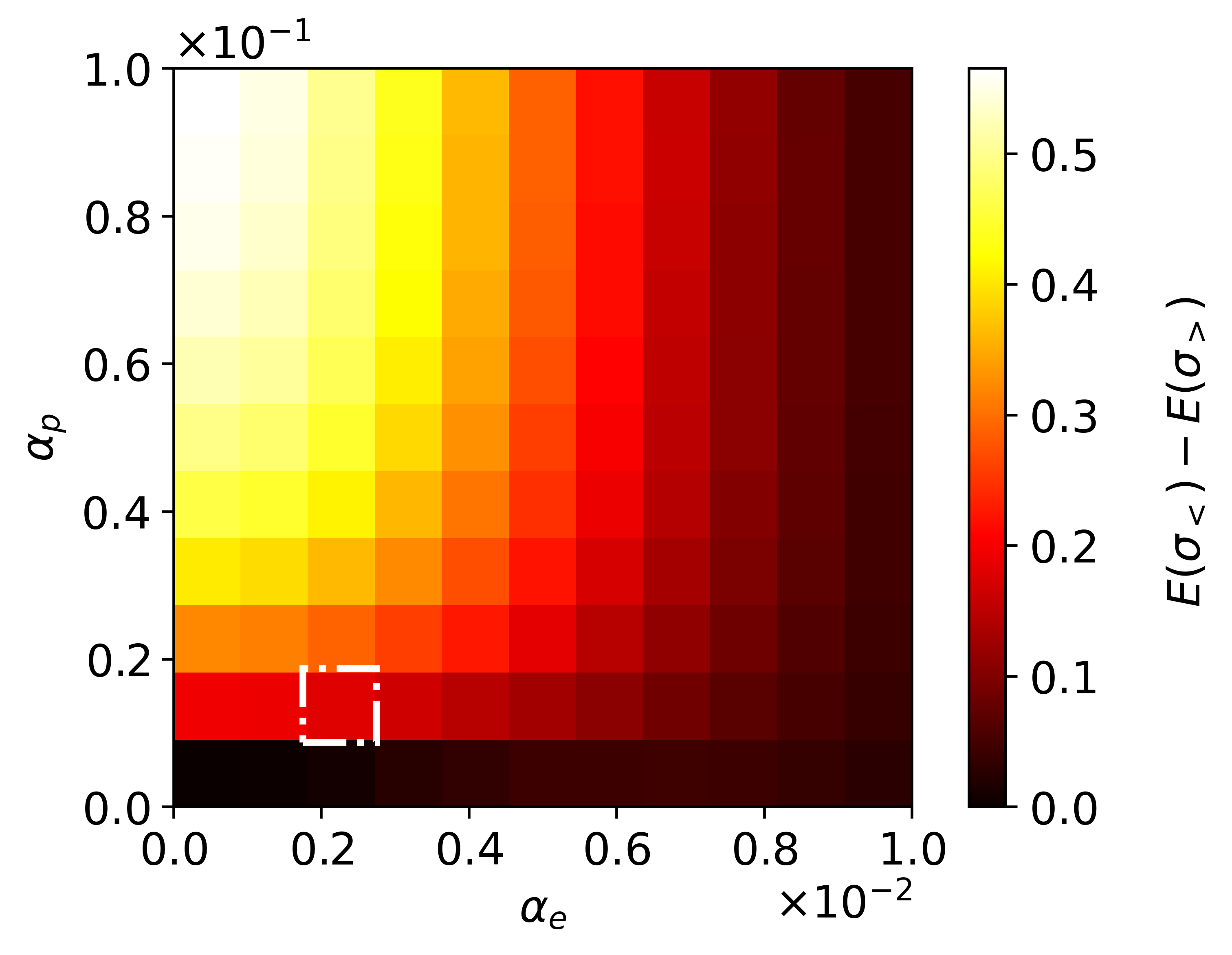}
			\caption{The difference in exposure risks of the two scenarios for social distancing intensity, varying $\alpha_p$ and $\alpha_e$ values.
			The color code indicates $E(\sigma_<) - E(\sigma_>)$.
			The dash-dotted pixel corresponds to the \textit{baseline} values of $\alpha_e = 2 \times 10^{-3}$ and $\alpha_p = 10^{-2}$.
			Please note the discrepancy in the order of magnitude of the axes.
			}
			\label{fig:two-sigmas-E-compare}
		\end{center}
	\end{figure}

	In Fig. \ref{fig:two-sigmas-E-compare} we compare the overall exposure risks for $\sigma_<$ and $\sigma_>$ scenarios. As shown in this figure, there is a significant discrepancy for a wide range of infection probabilities confirming our previous results on the effectiveness of social distancing.
	Due to the increasing role of direct infection and this type of infection's	sensitivity to social distancing, the discrepancy increases for higher $\alpha_p$ and lower $\alpha_e$ (upper left of the diagram).
	Please note that for high values of $\alpha_e$ (right) even when $\alpha_p$ is high (upper right), social distancing does not play a major role. This can be explained by the \textit{competition} between the direct and indirect transmission methods. In other words, in this area, even if agents lower the chance of direct transmission, following the social distancing guidelines, most of them will be nevertheless, exposed to the infection by the indirect transmission method.

\begin{figure*}[]
		\begin{center}
			\includegraphics[width=\linewidth]{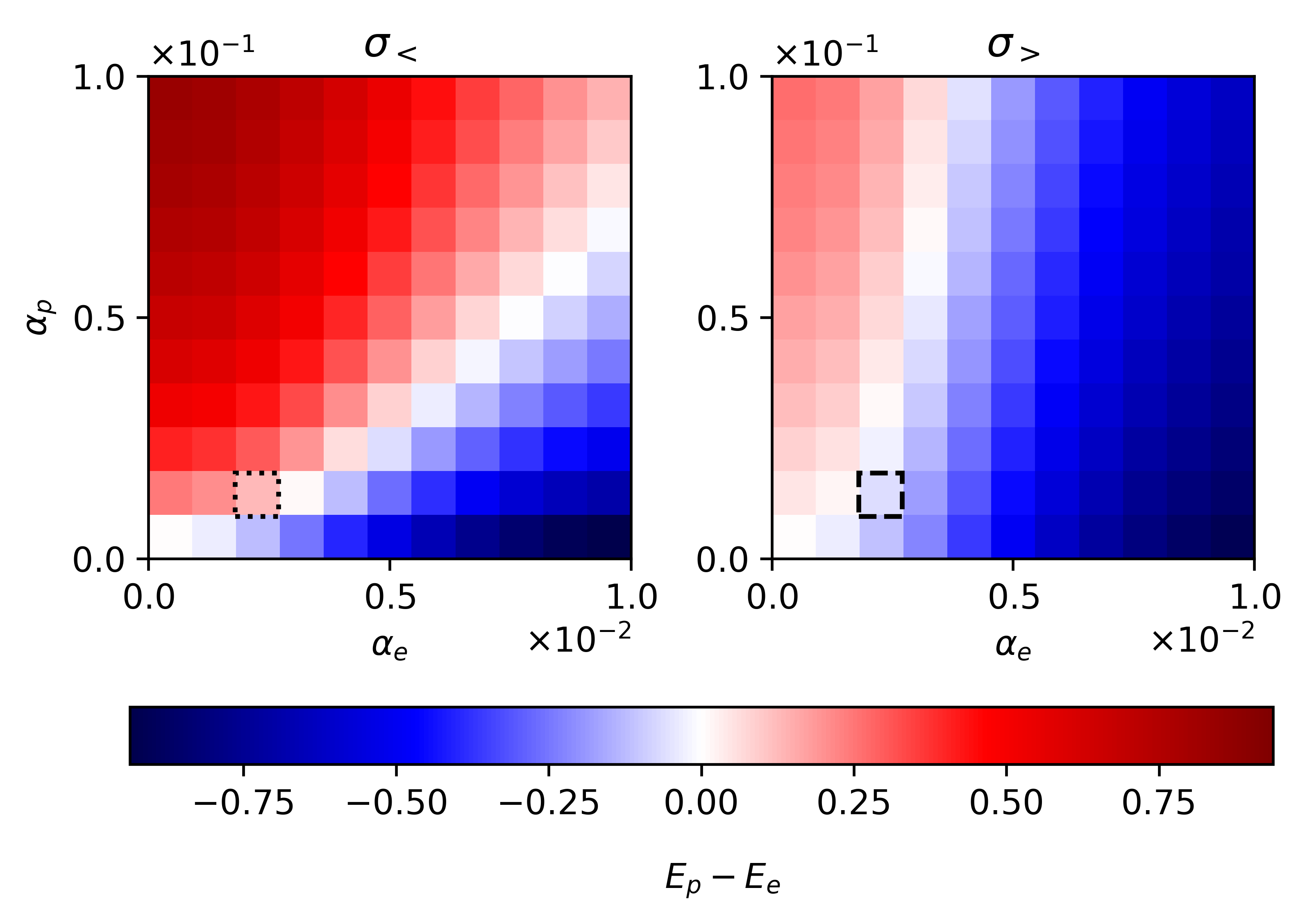}
			\caption{The effect of social distancing on contagion regimes. $E_p - E_e$ illustrated for $\alpha_p$ and $\alpha_e$ values.
			Left and right panels respectively denote $\sigma_<$ and $\sigma_>$ scenarios for social distancing intensity.
			The red and blue areas respectively demonstrate $E_p$ and $E_e$ dominated areas.
			The dotted and dashed pixels both correspond to the \textit{baseline} values of $\alpha_e = 2 \times 10^{-3}$ and $\alpha_p = 10^{-2}$.
			Please note the discrepancy in the order of magnitude of the axes.
			}
			\label{fig:high-sigma-deltaE}
		\end{center}
	\end{figure*}

	Furthermore, to investigate the role of direct and indirect transmission, we calculate $E_p - E_e$ (Fig. \ref{fig:high-sigma-deltaE}). We observe two regimes, $E_p$ and $E_e$ dominated areas, respectively denoted by red and blue and a white area illustrating the border between the regimes. By comparing the panels we see that due to social distancing, the $E_p$ dominated regime has drastically shrunk in favor of $E_e$ dominance.
	This observation also agrees with our previous findings that the person-to-person infection significantly decreases in the social distancing scenario $(\sigma_>)$ and environmental infection's role increases due to the higher number of available susceptible agents.
	
	\section{\label{disc}Summary and Discussion}
	We have modeled the spreading of infection among mobile agents,
	as a combination of pedestrian dynamics and compartmental spreading model ($SIE$). We defined social distancing as the intensity of social force in pedestrian dynamics, and mapped it to the average value of minimum distance between agents and their average speed and concentration.
	By taking into account both direct and indirect transmission methods of infection (Fig. \ref{fig:schematic-illustration}), we have systematically evaluated the agents' exposure risks for a wide range of spreading and mobility parameters.
	
	We observe that for social distancing to be \textit{executable}, the population density of the environment should be under certain values (Fig. \ref{fig:ideal-social-distancing}). As a side effect, by applying social distancing the speed of mobile agents and therefore their performance will decrease (Fig. \ref{fig:sigma_distance}). Although social distancing has drastic effect on hindering the direct transmission, some of its effects can be cancelled by the increase in the indirect transmission (Fig. \ref{fig:sigma_variaton_E}) and the \textit{effectiveness} of social distancing is dependent on the direct and indirect transmission probabilities (Fig. \ref{fig:two-sigmas-E-compare}).
	We demonstrated the direct transmission and indirect transmission dominated regimes in the scenarios of social distancing abidance and non-abidance (Fig. \ref{fig:high-sigma-deltaE}). We also studied the \textit{effectiveness} of social distancing when abided by a fraction of population (Fig. \ref{fig:two_type_sigma_variaton}), we found that even though the increase in abiding by social distancing reduces the risk of direct transmission, the measures when followed by the agents most likely to be infectious, have the greatest effect.
	
	According to our findings, social distancing does not always hinder the transmission, it may counter-intuitively enhance the transmission in specific regimes of $\sigma$ for contagions with relatively high indirect transmission. (consider Fig. \ref{fig:sigma_variaton_E} middle panel in the absence of direct transmission.) It is not clear whether this range of parameters conform to real world epidemic diseases and should be further investigated through experimental studies.
	
    Our findings can propose a guideline for policy-makers on how to decrease the exposure risk at public locations for many contagions such as SARS-CoV-2. The proposed model can be used to approximate the risk of exposure in a public environment where people exhibit pedestrian dynamics such as public transportation, religious sites, leisure centers, educational campuses, etc. For example, the management board of a shopping mall can determine the decrease needed in population density in order to reduce the risk of exposure by a desired factor. In the first place, the population density of an environment should be in the region where social distancing is \textit{executable}.
    Then the risk factor for this environment can be calculated via the proposed model for different populations. Now using these results, the corresponding population for a desired reduced risk factor can be derived. Afterwards to propose a specific value for $\sigma$, by considering the social distancing force's exposure risks and also its effect on the population performance, a desired $\sigma$ can be proposed for the shopping mall, the customers can be advised to keep a distance of $\braket{\overline{L}_{i,\mathcal{N}_{i}}}$ as calculated in
    Fig. \ref{fig:sigma_distance}
    for the corresponding value of $\sigma$. Also in order to keep social distancing and overcome the indirect transmission, periodic cleaning and disinfecting of the environment can be conducted. Cleaning the environment would be most important in contagions with higher $\alpha_e$ (See Fig. \ref{fig:two-sigmas-E-compare}).

	Despite the capabilities mentioned above, nevertheless, this model has some limitations worth noting. The infection probabilities $\alpha_p$ an $\alpha_e$, are hard to specify for an infection. Since they are the probability of getting infected in \textit{one time step} of the simulation, 
   	they should be specified based on the simulation time step size.
	It should be also noted that they are not purely biological parameters. For example, talking, shaking hands, and some other social behavior can alter the value of person to person infection probability $\alpha_p$. These social factors make the parameters harder to estimate due to the ethical and technical limitations in experiments.
	
    For further studies, the compartmental part of the model is versatile and can be expanded to simulate long term dynamics of the spreading by considering the $E$ to $I$ (exposed becoming infectious) and $I$ to $R$ (infectious becoming recovered) transitions. Also this model can be combined with other empirical or random generated networks, e.g. random geometric graphs \cite{barthelemy2011spatial,rodriguez2019particle,PhysRevE.66.016121} and mean-field models.
    
    Note that apart from the environmental infection, our model is theoretically equivalent to the spreading model on the temporal spatial network of interactions between the agents where there is a connection between agents closer than $r_s$. It is also possible to include environmental infection by considering tiles as another type of stationary nodes, reacting differently to the infection.

	\section*{Code Availability Statement}
    The simulation and analysis is conducted by \textit{Episterian} software written in Python, developed by S.S and A.H, available
        	\href{https://github.com/ialireza13/Episterian}{\textbf{\textcolor{blue}{here}}}
        	under GPLv3.
	
	\section*{Acknowledgments}
	We would like to thank M. R. Ejtehadi for his insightful comments. F.Gh.
acknowledges partial support by Deutsche Forschungs-gemeinschaft (DFG) under the grant (idonate project:345463468).
	
    	\section*{Supporting Information}
    	\subsection{Baseline parameters values}
    	The dashed, dotted and dash-dotted lines and areas illustrated in Figs.
    	\ref{fig:sigma_distance},
    	\ref{fig:sigma_variaton_E},
    	\ref{fig:two_type_sigma_variaton},
    	\ref{fig:two-sigmas-E-compare} and
    	\ref{fig:high-sigma-deltaE}
    	correspond to the baseline values specified in table \ref{tbl:basline-values}.
    	    	
    	\begin{table}[ht]
		\caption{Baseline parameters values}
		\begin{tabular*}{0.48\textwidth}{c @{\extracolsep{\fill}} ccccc}
		Sign & Schematic    & $\sigma$   & $\alpha_p$ &  $\alpha_e$  \\ \hline
		dotted
		
		&  $\cdot\,\cdot\,\cdot$
		& $\sigma_< = 0.3$
		&$10^{-2}$
		& $2 \times 10^{-3}$
		\\ \hline
		
		dashed
		& \rule[.5ex]{0.5em}{.4pt}\,\rule[.5ex]{0.5em}{.4pt}\,\rule[.5ex]{0.5em}{.4pt}
		& $\sigma_> = 1.5$
		&$10^{-2}$
		& $2 \times 10^{-3}$
		\\ \hline
		
		dash-dotted
		& \rule[.5ex]{0.5em}{.4pt}\,$\cdot$\,\rule[.5ex]{0.5em}{.4pt}
		& N.A
		&$10^{-2}$
		& $2 \times 10^{-3}$
		\\
		\label{tbl:basline-values}
		\end{tabular*}
		\end{table}

	\clearpage	
\bibliographystyle{ieeetr}

\begin{thebibliography}{10}
	
	\bibitem{noauthor_social_nodate}
	CDC-NCIRD-DVD, ``Social {Distancing}, {Quarantine}, and {Isolation}.''
	
	\bibitem{kermack_mckendrick_1927}
	W.~O. Kermack and A.~G. Mckendrick, ``A contribution to the mathematical theory
	of epidemics,'' {\em Proceedings of the Royal Society A: Mathematical,
		Physical and Engineering Sciences}, vol.~115, p.~700–721, Jan 1927.
	
	\bibitem{keeling2011modeling}
	M.~J. Keeling and P.~Rohani, {\em Modeling infectious diseases in humans and
		animals}.
	\newblock Princeton University Press, 2011.
	
	\bibitem{newman_2018}
	M.~E.~J. Newman, {\em Networks an introduction}.
	\newblock Oxford University Press, 2018.
	
	\bibitem{barrat}
	A.~Barrat, M.~Barthelemy, and A.~Vespignani, {\em Dynamical processes on
		complex networks}.
	\newblock Cambridge University Press, 2008.
	
	\bibitem{masuda2017temporal}
	N.~Masuda and P.~Holme, {\em Temporal network epidemiology}.
	\newblock Springer, 2017.
	
	\bibitem{temporalnetworksholme}
	P.~Holme and S.~Jari, {\em Temporal Networks}.
	\newblock Springer, 2013.
	
	\bibitem{smallbutslowworld}
	M.~Karsai, M.~Kivelä, R.~K. Pan, K.~Kaski, J.~Kertész, A.~L. Barabási, and
	J.~Saramäki, ``Small but slow world: How network topology and burstiness
	slow down spreading,'' 2010.
	
	\bibitem{fakhteh-hospital}
	J.~P. Rodríguez, F.~Ghanbarnejad, and V.~M. Eguíluz, ``Risk of coinfection
	outbreaks in temporal networks: A case study of a hospital contact network,''
	{\em Frontiers in Physics}, vol.~5, Jun 2017.
	
	\bibitem{sajjadi_impact_2020}
	S.~Sajjadi, M.~R. Ejtehadi, and F.~Ghanbarnejad, ``Impact of temporal
	correlations on high risk outbreaks of independent and cooperative {SIR}
	dynamics,'' {\em arXiv:2003.01268 [physics, q-bio]}, Mar. 2020.
	\newblock arXiv: 2003.01268.
	
	\bibitem{helbing_social_1995}
	D.~Helbing and P.~Molnar, ``Social {Force} {Model} for {Pedestrian}
	{Dynamics},'' {\em Physical Review E}, vol.~51, pp.~4282--4286, May 1995.
	\newblock arXiv: cond-mat/9805244.
	
	\bibitem{namilae_self-propelled_2017}
	S.~Namilae, A.~Srinivasan, A.~Mubayi, M.~Scotch, and R.~Pahle, ``Self-propelled
	pedestrian dynamics model: {Application} to passenger movement and infection
	propagation in airplanes,'' {\em Physica A: Statistical Mechanics and its
		Applications}, vol.~465, pp.~248--260, Jan. 2017.
	
	\bibitem{harweg_agent-based_2020}
	T.~Harweg, D.~Bachmann, and F.~Weichert, ``Agent-based {Simulation} of
	{Pedestrian} {Dynamics} for {Exposure} {Time} {Estimation} in {Epidemic}
	{Risk} {Assessment},'' {\em arXiv:2007.04138 [physics]}, July 2020.
	\newblock arXiv: 2007.04138.
	
	\bibitem{namilae_multiscale_2017}
	S.~Namilae, P.~Derjany, A.~Mubayi, M.~Scotch, and A.~Srinivasan, ``Multiscale
	model for pedestrian and infection dynamics during air travel,'' {\em
		Physical Review E}, vol.~95, p.~052320, May 2017.
	
	\bibitem{kim_coupling_2020}
	D.~Kim and A.~Quaini, ``Coupling kinetic theory approaches for pedestrian
	dynamics and disease contagion in a confined environment,'' {\em
		arXiv:2003.08357 [physics, q-bio]}, Apr. 2020.
	\newblock arXiv: 2003.08357.
	
	\bibitem{gosce_analytical_2015}
	L.~Goscé, D.~A.~W. Barton, and A.~Johansson, ``Analytical {Modelling} of the
	{Spread} of {Disease} in {Confined} and {Crowded} {Spaces},'' {\em Scientific
		Reports}, vol.~4, p.~4856, May 2015.
	
	\bibitem{xiao_modeling_nodate}
	Y.~Xiao and M.~Yang, ``Modeling indoor-level non-pharmaceutical interventions
	during the covid-19 pandemic: a pedestrian dynamics-based microscopic
	simulation approach,'' p.~24.
	
	\bibitem{bouchnita_multi-scale_2020}
	A.~Bouchnita and A.~Jebrane, ``A multi-scale model quantifies the impact of
	limited movement of the population and mandatory wearing of face masks in
	containing the {COVID}-19 epidemic in {Morocco},'' {\em Mathematical
		Modelling of Natural Phenomena}, vol.~15, p.~31, 2020.
	
	\bibitem{derjany_multiscale_2020}
	P.~Derjany, S.~Namilae, D.~Liu, and A.~Srinivasan, ``Multiscale model for the
	optimal design of pedestrian queues to mitigate infectious disease spread,''
	{\em PLOS ONE}, vol.~15, p.~e0235891, July 2020.
	
	\bibitem{us2013principles}
	U.~D. of~Health, H.~Services, {\em et~al.}, ``Principles of epidemiology in
	public health practice third edition an introduction to applied epidemiology
	and biostatistics,'' {\em Atlanta, Georgia, USA Available on the website:
		https://www.cdc.gov/csels/dsepd/ss1978/SS1978.pdf}, vol.~8, 2013.
	
	\bibitem{castellano_statistical_2009}
	C.~Castellano, S.~Fortunato, and V.~Loreto, ``Statistical physics of social
	dynamics,'' {\em Reviews of Modern Physics}, vol.~81, pp.~591--646, May 2009.
	\newblock arXiv: 0710.3256.
	
	\bibitem{wiki:euler-method}
	{Wikipedia contributors}, ``Euler method --- {Wikipedia}{,} the free
	encyclopedia,'' 2020.
	\newblock [Online; accessed 22-August-2020].
	
	\bibitem{wiki:incubation}
	{Wikipedia contributors}, ``Incubation period --- {Wikipedia}{,} the free
	encyclopedia.''
	\url{https://en.wikipedia.org/w/index.php?title=Incubation_period&oldid=973832885},
	2020.
	\newblock [Online; accessed 10-September-2020].
	
	\bibitem{rejection-based}
	C.~L. Vestergaard and M.~Génois, ``Temporal gillespie algorithm: Fast
	simulation of contagion processes on time-varying networks,'' {\em PLOS
		Computational Biology}, vol.~11, no.~10, 2015.
	
	\bibitem{onemeter}
	{World health organization}, ``Coronavirus disease (covid-19) advice for the
	public.''
	\url{https://www.who.int/emergencies/diseases/novel-coronavirus-2019/advice-for-public},
	2020.
	\newblock [Online; accessed 10-September-2020].
	
	\bibitem{barthelemy2011spatial}
	M.~Barth{\'e}lemy, ``Spatial networks,'' {\em Physics Reports}, vol.~499,
	no.~1-3, pp.~1--101, 2011.
	
	\bibitem{rodriguez2019particle}
	J.~P. Rodr{\'\i}guez, F.~Ghanbarnejad, and V.~M. Egu{\'\i}luz, ``Particle
	velocity controls phase transitions in contagion dynamics,'' {\em Scientific
		reports}, vol.~9, no.~1, pp.~1--9, 2019.
	
	\bibitem{PhysRevE.66.016121}
	J.~Dall and M.~Christensen, ``Random geometric graphs,'' {\em Phys. Rev. E},
	vol.~66, p.~016121, Jul 2002.
	
\end{thebibliography}

\end{document}